\documentclass[letterpaper,10pt,conference,twocolumn]{IEEEtran}

\IEEEoverridecommandlockouts

\usepackage[english]{babel}
\usepackage[latin1]{inputenc}
\usepackage[T1]{fontenc}
\usepackage{url}
\usepackage{graphicx}
\usepackage{subfigure}
\usepackage{color}
\usepackage[cmex10]{amsmath}
\usepackage{amsthm,amsfonts,amssymb}
\usepackage{colortbl}                   
\usepackage{url}
\usepackage{booktabs}
\usepackage{epsfig}
\usepackage{pstricks}
\usepackage{pst-node}
\usepackage{pst-grad}
\usepackage{pst-plot}
\usepackage{pstricks-add}
\usepackage{ifthen}
\usepackage{algorithm}
\usepackage{algpseudocode}
\usepackage{float}
\usepackage{fp}
\usepackage{cite}

\newfloat{algorithm}{t}{lop}

\newboolean{draft}
\newcommand{\isdraft}[2]{\ifthenelse{\boolean{draft}}{#1}{#2}}

\isdraft{\usepackage{setspace}}{}                              
\isdraft{\usepackage[footnotesize]{caption}}{}                 
\isdraft{\usepackage{paralist}}{}                              

\def\today{\ifcase\month\or
  January\or February\or March\or April\or May\or June\or
  July\or August\or September\or October\or November\or December\fi \space \number\year}

\newcommand{\fref}[1]{Fig.~\ref{#1}}

\newcommand{\cref}[1]{Chapter~\ref{#1}}

\newcommand{\pref}[1]{Proposition~\ref{#1}}
\newcommand{\thref}[1]{Theorem~\ref{#1}}

\newcommand{\Arxiv}[1]{preprint: \url{#1}}
\newcommand{\OptimizationOnline}[1]{preprint: \url{#1}}

\theoremstyle{plain}
\newtheorem{Definition}{Definition}
\newtheorem{Theorem}{Theorem}

\newtheorem{Proposition}[Theorem]{Proposition}

\newcommand{\mypar}[1]{{\bf #1.}}

\isdraft{}{}

\title{Compressed Sensing With Side Information: \\	
	Geometrical Interpretation and Performance Bounds
}

\begin{document}

\author{\IEEEauthorblockN{Jo\~ao F.~C.~Mota, Nikos Deligiannis, and Miguel R.~D.~Rodrigues}
\IEEEauthorblockA{Electronic \& Electrical Engineering Department at University College London, UK}
	\thanks{Work funded by EPSRC grant EP/K033166/1.}
}

\maketitle

\begin{abstract}
	We address the problem of Compressed Sensing (CS) with side information. Namely, when reconstructing a target CS signal, we assume access to a similar signal. This additional knowledge, the \textit{side information}, is integrated into CS via $\ell_1$-$\ell_1$ and $\ell_1$-$\ell_2$ minimization. We then provide lower bounds on the number of measurements that these problems require for successful reconstruction of the target signal. If the side information has good quality, the number of measurements is significantly reduced via $\ell_1$-$\ell_1$ minimization, but not so much via $\ell_1$-$\ell_2$ minimization. We provide geometrical interpretations and experimental results illustrating our findings. 
\end{abstract}


\section{Introduction}

	Consider an unknown signal~$x^\star \in \mathbb{R}^n$ that is $s$-sparse, i.e., at most~$s$ entries are nonzero.	
	Assume we take~$m<n$ linear measurements of~$x^\star$: $y=A x^\star$, where~$A \in \mathbb{R}^{m\times n}$. Compressed Sensing (CS)~\cite{Donoho06-CompressedSensing} answers two fundamental questions: \textit{How to reconstruct~$x^\star$ from~$y$? And how many measurements~$m$ do we need for successful reconstruction?} If~$A$ satisfies the RIP~\cite{Candes05-DecodingByLinearProgramming} or a null-space property~\cite{Chandrasekaran12-ConvexGeometryLinearInverseProblems}, CS establishes that~$x^\star$ can be reconstructed by solving \textit{basis pursuit}~\cite{Donoho98-AtomicDecompositionBasisPursuit}:
	\begin{equation}\label{Eq:BP}
		\begin{array}[t]{ll}
			\underset{x}{\text{minimize}} & \|x\|_1 \\
			\text{subject to} & Ax=y\,.
		\end{array}
	\end{equation}
	In particular, when the entries of~$A$ are i.i.d.\ Gaussian, then~$m > 2s\log(n/s) + (7/5)s$ measurements guarantee that~$x^\star$ is the unique solution of~\eqref{Eq:BP} with high probability~\cite{Chandrasekaran12-ConvexGeometryLinearInverseProblems}, a bound known to be tight for Gaussian matrices~\cite{Tropp13-LivingOnTheEdge}.

	\mypar{CS with side information}
	Suppose we have access to \textit{side information}, a vector~$w \in \mathbb{R}^n$ that is similar to~$x^\star$. This occurs when reconstructing sequences of signals (e.g., video~\cite{Kang09-DistributedCompressiveVideoSensing} and estimation problems~\cite{Charles11-SparsityPenaltiesDynamicalSystemEstimation}), or when we have access to prior similar signals (e.g., sensor networks~\cite{Baron06-DistributedCompressedSensing}, multiview cameras~\cite{Trocan10-DisparityCompensatedCompressedSensingMultiviewImages}, and medical imaging~\cite{Chen08-PriorImageConstrainedCS}). Our goal is to answer the same questions that CS does, but with the additional knowledge of side information: \textit{How to reconstruct~$x^\star$  from the measurements~$y$ and the side information~$w$? And how many measurements~$m$ do we need for successful reconstruction?}
	
\mypar{Our strategy} 
	Let~$g:\mathbb{R}^n \xrightarrow{} \mathbb{R}$ be a function that models similarity between~$w$ and~$x^\star$: the smaller~$g(x^\star - w)$, the higher the similarity. A natural approach to integrate~$w$ into~\eqref{Eq:BP} is to
	\begin{equation}\label{Eq:BPSideInfoGeneric}
		\begin{array}[t]{ll}
			\underset{x}{\text{minimize}} & \|x\|_1 + \beta\, g(x-w) \\
			\text{subject to} &  Ax=y\,,
		\end{array}
	\end{equation}
	where~$\beta > 0$. We consider two models for~$g$: $g_1(\cdot) := \|\cdot\|_1$ and~$g_2(\cdot) = (1/2)\|\cdot\|_2^2$; and refer to~\eqref{Eq:BPSideInfoGeneric} with~$g = g_1$ as $\ell_1$-$\ell_1$ minimization and to~\eqref{Eq:BPSideInfoGeneric} with~$g = g_2$ as $\ell_1$-$\ell_2$ minimization. Although instances and variations of~\eqref{Eq:BPSideInfoGeneric} with~$g_1$ and~$g_2$ have appeared in the literature (see Related work below), to the best of our knowledge, no CS-like recovery guarantees have ever been provided. 
	
	Assuming the entries of~$A$ are i.i.d.\ Gaussian, we compute bounds on the number of measurements above which $\ell_1$-$\ell_1$ and $\ell_1$-$\ell_2$ minimization reconstruct~$x^\star$ perfectly, with high probability. When the side information is ``good enough,'' our bound for $\ell_1$-$\ell_1$ minimization is much smaller than the bounds both for $\ell_1$-$\ell_2$ minimization and for classical CS. In addition, our experiments confirm that $\ell_1$-$\ell_1$ minimization requires in general less measurements for successful reconstruction than both $\ell_1$-$\ell_2$ minimization and classical CS. We explain this phenomenon using the underlying geometry of the problem. Proofs of the results presented in this paper can be found in~\cite{Mota14-CompressedSensingSideInformation}. For succinctness, we consider here only the case~$\beta = 1$ in~\eqref{Eq:BPSideInfoGeneric}, but results for~$\beta \neq 1$ can be found in~\cite{Mota14-CompressedSensingSideInformation}. 

\mypar{Related work}	
	Several methods improve the performance of CS by assuming access to side (or prior) information. The majority, however, uses concepts of side information different from ours, for example, estimates on the support of~$x^\star$~\cite{Vaswani10-ModifiedCS}, or its probability distribution~\cite{Scarlett13-CompressedSensingPriorInformation}. The first work using side information in our sense, namely $\ell_1$-$\ell_1$ minimization, appears to be~\cite{Chen08-PriorImageConstrainedCS}. That work focuses on the application of computed tomography and does not provide either any type of analysis or a comparison with standard CS; see~\cite{Eldar14-ApplicationCSLongitudinalMRI} for a recent related approach. In~\cite{Vaswani10-ModifiedCS}, a problem similar to $\ell_1$-$\ell_2$ minimization appears as an extension of the main problem studied in that paper. Although experimental results are presented, no analysis is provided for the $\ell_1$-$\ell_2$-type problem. Prior work also has considered the Lagrangian version of~\eqref{Eq:BPSideInfoGeneric} where there are no constraints, but the extra term~$\lambda\|y - Ax\|_2^2$ is added to the objective, with~$\lambda > 0$. For example, \cite{Charles11-SparsityPenaltiesDynamicalSystemEstimation} estimates the state of a dynamical system using the previous instant's state as side information. The estimation problem is posed as the Lagrangian version of~\eqref{Eq:BPSideInfoGeneric} with both~$g_1$ and~$g_2$. Although the experimental results in~\cite{Charles11-SparsityPenaltiesDynamicalSystemEstimation} indicate that $\ell_1$-$\ell_1$-type of minimization requires less measurements than $\ell_1$-$\ell_2$, no rationale is given. Our theoretical results and geometrical interpretations explain this phenomenon in the context of~\eqref{Eq:BPSideInfoGeneric}. 
	Finally, the work in~\cite{Wang13-SideInformationAidedCS} analyzes the performance of a message passing algorithm to solve the Lagrangian version of~\eqref{Eq:BPSideInfoGeneric} with~$g_2$. 


\section{Definitions And Geometrical Interpretations}

	As mentioned before, \cite{Chandrasekaran12-ConvexGeometryLinearInverseProblems} establishes tight bounds for CS. The main tool is the concept of \textit{Gaussian width of a cone}~$C \subset \mathbb{R}^n$, given by $w(C) := \mathbb{E}_g \bigl[\,\sup_z\{g^\top z\,:\, z \in C \cap B_n(0,1)\}\,\bigr]$,
	where~$g \in \mathbb{R}^n$ has i.i.d.\ zero-mean, unit variance Gaussian entries, and~$\mathbb{E}_g[\cdot]$ is the expected value w.r.t.\ $g$. We use~$B_n(0,1):= \{x \in \mathbb{R}^n\,:\, \|x\|_2 \leq 1\}$ to denote the unit $\ell_2$-norm ball in~$\mathbb{R}^n$. The Gaussian width was originally proposed in~\cite{Gordon88-EscapeThroughTheMesh} for measuring the ``width'' (aperture) of a cone. Related work using this concept includes~\cite{Rudelson08-SparseReconstructionFourierGaussianMeasurements,Stojnic09-VariousThresholdsForL1OptimizationInCompressedSensing,Rao12-UniversalMeasurementBoundsForStructuredSparseSignalRecovery,Tropp13-LivingOnTheEdge,Oymak13-SquaredErrorGeneralizedLASSO,Chandrasekaran13-ComputationalStatisticalTradeoffsViaConvexRelaxation,Oymak13-SharpMSEBoundsProximalDenoising,Oymak13-SimultaneousStructuredModelsSparseLowRankMatrices,Foygel14-CorruptedSensing,Bandeira14-CompressiveClassificationRareEclipseProblem,Tropp14-ConvexRecoveryStructuredSignalFromIndependentRandomLinearMeasurements,Kabanava14-RobustAnalysisL1RecoveryGaussianMeasurementsTV}.
	\begin{Theorem}[Corollary 3.3 in \cite{Chandrasekaran12-ConvexGeometryLinearInverseProblems}]
	\label{Thm:Chandrasekaran}
		Let~$A \in \mathbb{R}^{m\times n}$ be a matrix whose entries are i.i.d.\ zero-mean Gaussian random variables with variance~$1/m$, and let~$f:\mathbb{R}^n \xrightarrow{} \mathbb{R}$ be a convex function. Assume~$y = Ax^\star$ and that $m \geq w(T_f(x^\star))^2 + 1$, where~$T_f(x^\star)$ denotes the tangent cone of~$f$ at~$x^\star$. Consider	
		\begin{equation}\label{Eq:ThmChandrasekaranNoiseless}
			\hat{x} \in 
			\begin{array}[t]{cl}
				\underset{x}{\arg\min} & f(x) \\
				\text{\emph{s.t.}} &  A x = y\,.
			\end{array}
		\end{equation}
		Then, $\hat{x} = x^\star$ is the unique solution of~\eqref{Eq:ThmChandrasekaranNoiseless} with probability at least $1-\exp\bigl(-\frac{1}{2}\bigl[ w(T_f(x^\star)) - \lambda_m\bigr]^2\bigr)$, where~$\lambda_m$ is the expected length of a zero-mean, unit-variance Gaussian vector in~$\mathbb{R}^m$.
	\end{Theorem}
	Recall that the tangent cone of a convex function~$f$ at a point~$x^\star$ is $T_f(x^\star) := \text{cone} \{x - x^\star\,:\, f(x) \leq f(x^\star)\}$, where~$\text{cone}\,C := \{\alpha c\,:\, \alpha \geq 0,\, c \in C\}$ is the cone generated by the set~$C$. In other words, $T_f(x^\star)$ is the cone generated by the \textit{sublevel set} $S_f(x^\star) := \{x \in \mathbb{R}^n\,:\, f(x) \leq f(x^\star)\}$ from the point~$x^\star$, i.e., $T_f(x^\star) = \text{cone}(S_f(x^\star) - x^\star)$. Note that~\eqref{Eq:ThmChandrasekaranNoiseless} becomes~\eqref{Eq:BP} when~$f(x) = \|x\|_1$ and becomes~\eqref{Eq:BPSideInfoGeneric} when~$f(x) = \|x\|_1 + g(x,w)$. Note also that $m/\sqrt{m+1} \leq \lambda_m \leq \sqrt{m}$~\cite{Chandrasekaran12-ConvexGeometryLinearInverseProblems}. \thref{Thm:Chandrasekaran} states that~\eqref{Eq:ThmChandrasekaranNoiseless} recovers~$x^\star$ with high probability if the number of measurements is larger than the squared Gaussian width of the tangent cone of~$f$ at~$x^\star$. The work in~\cite{Chandrasekaran12-ConvexGeometryLinearInverseProblems} then establishes:
	\begin{Proposition}[Proposition 3.10 in~\cite{Chandrasekaran12-ConvexGeometryLinearInverseProblems}]
	\label{Prop:ChandrasekaranBound}
		Let~$x^\star \neq 0$ be an $s$-sparse vector in~$\mathbb{R}^n$. Then, 
		\begin{equation}\label{Eq:PropChandrasekaranBound}
			w\bigl(T_{\|\cdot\|_1}(x^\star)\bigr)^2 \leq 2s\log\Bigl(\frac{n}{s}\Bigr) + \frac{7}{5}s\,.
		\end{equation} 
	\end{Proposition}
	By upper bounding the squared Gaussian width of~$T_{\|\cdot\|_1}(x^\star)$, \pref{Prop:ChandrasekaranBound} establishes a lower bound on the number of measurements that~\eqref{Eq:BP} requires to recover~$x^\star$ with high probability. 
	Note that, since~$w(T_f(x^\star))$ is usually unknown, \thref{Thm:Chandrasekaran} is not very informative in practice. \pref{Prop:ChandrasekaranBound} instills it with operational significance by upper bounding~$w(T_f(x^\star))^2$ in terms of the key signal parameters~$s$ and~$n$. Our goal is to do the same for the functions~$f_1(x) := \|x\|_1 + \|x - w\|_1$ and~$f_2(x) := \|x\|_1 + \frac{1}{2}\|x - w\|_2^2$.

	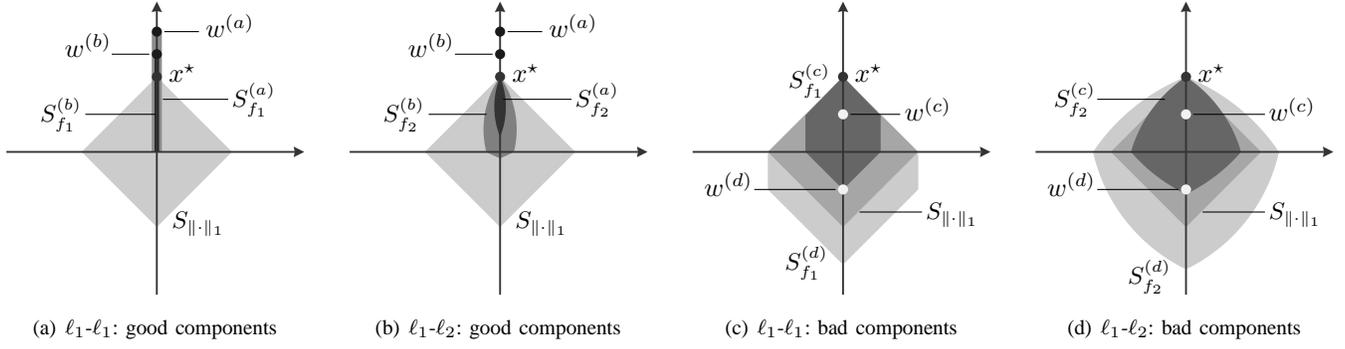
\begin{figure*}
	\centering	
		
	\subfigure[$\ell_1$-$\ell_1$: good components]{\label{SubFig:IllustrationL1L1Good}
	\psscalebox{1.0}{
	\begin{pspicture}(3.96,3.9)
		\psset{blendmode=2}	
	
	\def\radius{1}       
								
	\def\diamond{       
		\pspolygon*[linecolor=black!20!white](0,\radius)(\radius,0.0)(0.0,-\radius)(-\radius,0.0)	
	}

	\rput(2.0,2.0){\diamond}        
											
	\psset{linecolor=black!45!white}
	\psline[linewidth=3.8pt](2.0,2.0)(2.0,3.6)
	
	\psset{linecolor=black!85!white}
	\psline[linewidth=1.7pt](2.0,2.0)(2.0,3.3)

	\psset{linestyle=solid,linecolor=black!80!white,arrowsize=4pt,arrowinset=0.1,arrowlength=1.0}
	
  \psline[linewidth=0.7pt]{->}(0.0,2.0)(3.96,2.0)
  \psline[linewidth=0.7pt]{->}(2.0,0.1)(2.0,4.0)
		
	\rput(2.0,3.3){\pscircle*[linecolor=black!90!white](0,0){0.067}}
	\rput(2.0,3.6){\pscircle*[linecolor=black!90!white](0,0){0.067}}	
			
	\rput(2.0,3.0){\rnode{x}{\pscircle*(0,0){0.067}}}
	
	
	\psset{linestyle=solid,linecolor=black,linewidth=0.4pt}
	
	\rput[l](2.16,3.05){$x^\star$}
	
	\rput[l](2.65,3.68){$w^{(a)}$}
	\psline{-}(2.1,3.6)(2.56,3.6)
	
	\rput[r](1.4,3.42){$w^{(b)}$}
	\psline{-}(1.9,3.3)(1.4,3.3)
		
	\rput[l](3.0,2.7){\small $S_{f_1}^{(a)}$}
	\psline{-}(2.08,2.7)(2.95,2.7)
	\rput[r](1.0,2.5){\small $S_{f_1}^{(b)}$}	
	\psline{-}(1.05,2.5)(1.98,2.5)
	\rput[tl](2.20,1.20){\small $S_{\|\cdot\|_1}$}
	
	\end{pspicture}
	}
	}
	\subfigure[$\ell_1$-$\ell_2$: good components]{\label{SubFig:IllustrationL1L2Good}
	\psscalebox{1.0}{
	\begin{pspicture}(3.96,3.9)
		\psset{blendmode=2}	
	
	\def\radius{1}       
								
	\def\diamond{       
		\pspolygon*[linecolor=black!20!white](0,\radius)(\radius,0.0)(0.0,-\radius)(-\radius,0.0)	
	}

	\rput(2.0,2.0){\diamond}

	\psset{linecolor=black!50!white}
	\pspolygon*
	(1.923065,2.874740)
  (1.901287,2.831831)
  (1.880715,2.787145)
  (1.861901,2.741586)
  (1.845324,2.696098)
  (1.830520,2.649861)
  (1.817511,2.602927)
  (1.806320,2.555350)
  (1.796967,2.507185)
  (1.789470,2.458485)
  (1.784190,2.410246)
  (1.780763,2.361596)
  (1.779200,2.312591)
  (1.779803,2.264244)
  (1.782256,2.215664)
  (1.786569,2.166905)
  (1.792750,2.118023)
  (1.801014,2.070051)
  (1.811118,2.022074)
  (1.826634,1.994841)
  (1.845718,1.984655)
  (1.865142,1.974832)
  (1.884897,1.965383)
  (1.904976,1.956317)
  (1.925302,1.946645)
  (1.945976,1.937372)
  (1.966992,1.928508)
  (1.988329,1.919063)
  (2.009958,1.918046)
  (2.031331,1.927458)
  (2.052381,1.936289)
  (2.073021,1.946528)
  (2.093451,1.955171)
  (2.113455,1.965201)
  (2.133225,1.974619)
  (2.152664,1.984410)
  (2.171934,1.993581)
  (2.187889,2.018828)
  (2.198335,2.065822)
  (2.206527,2.114773)
  (2.213090,2.162690)
  (2.217558,2.211462)
  (2.220166,2.260060)
  (2.220922,2.308430)
  (2.219514,2.357462)
  (2.216239,2.406144)
  (2.210750,2.455353)
  (2.203766,2.503161)
  (2.194563,2.551373)
  (2.183521,2.599000)
  (2.170660,2.645988)
  (2.156001,2.692285)
  (2.139097,2.738719)
  (2.120896,2.783463)
  (2.100464,2.828221)
  (2.078823,2.871206)
  (2.054974,2.914082)
  (2.000000,3.000000)

	\psset{linecolor=black!80!white}
	\pspolygon*
	(1.964922,2.897855)
  (1.953822,2.856234)
  (1.944664,2.815103)
  (1.936870,2.772600)
  (1.930759,2.729728)
  (1.926342,2.686538)
  (1.923837,2.644057)
  (1.922816,2.600386)
  (1.923685,2.557533)
  (1.926239,2.514572)
  (1.930484,2.471553)
  (1.936533,2.429520)
  (1.944146,2.386537)
  (1.953526,2.344646)
  (1.964559,2.302900)
  (1.977245,2.261350)
  (1.991590,2.221045)
  (2.007197,2.218033)
  (2.021666,2.258316)
  (2.034478,2.299848)
  (2.045638,2.341580)
  (2.055144,2.383461)
  (2.062994,2.425443)
  (2.069061,2.468468)
  (2.073433,2.511488)
  (2.076116,2.554455)
  (2.077112,2.597317)
  (2.076427,2.640024)
  (2.073841,2.683500)
  (2.069797,2.725742)
  (2.063831,2.768644)
  (2.056182,2.811181)
  (2.046862,2.853303)
  (2.035885,2.894961)
  (2.000000,3.000000)
		
	\psset{linestyle=solid,linecolor=black!80!white,arrowsize=4pt,arrowinset=0.1,arrowlength=1.0}
	
  \psline[linewidth=0.7pt]{->}(0.0,2.0)(3.96,2.0)
  \psline[linewidth=0.7pt]{->}(2.0,0.1)(2.0,4.0)
		
	\rput(2.0,3.3){\pscircle*[linecolor=black!90!white](0,0){0.067}}
	\rput(2.0,3.6){\pscircle*[linecolor=black!90!white](0,0){0.067}}	
			
	\rput(2.0,3.0){\rnode{x}{\pscircle*(0,0){0.067}}}
	
	
	\psset{linestyle=solid,linecolor=black,linewidth=0.4pt}
	
	\rput[l](2.16,3.05){$x^\star$}
	
	\rput[l](2.65,3.68){$w^{(a)}$}
	\psline{-}(2.1,3.6)(2.56,3.6)
	
	\rput[r](1.4,3.42){$w^{(b)}$}
	\psline{-}(1.9,3.3)(1.4,3.3)
		
	\rput[l](3.0,2.7){\small $S_{f_2}^{(a)}$}
	\psline{-}(2.04,2.7)(2.95,2.7)
	\rput[r](1.0,2.5){\small $S_{f_2}^{(b)}$}	
	\psline{-}(1.05,2.5)(1.82,2.5)
	\rput[tl](2.20,1.20){\small $S_{\|\cdot\|_1}$}
	
	\end{pspicture}
	}
	}
	\subfigure[$\ell_1$-$\ell_1$: bad components]{\label{SubFig:IllustrationL1L1Bad}
	\psscalebox{1.0}{
	\begin{pspicture}(3.96,3.9)
		\psset{blendmode=2}	
	
	\def\radius{1}       
								
	\def\diamond{       
		\pspolygon*[linecolor=black!35!white](0,\radius)(\radius,0.0)(0.0,-\radius)(-\radius,0.0)	
	}

	\psset{linecolor=black!20!white}
	\pspolygon*	
	(0.999529,1.969877)
  (0.999864,1.928140)
  (0.999473,1.883780)
  (0.999813,1.838227)
  (0.999673,1.789940)
  (0.999830,1.739627)
  (0.999791,1.686462)
  (0.999720,1.630381)
  (0.999781,1.571330)
  (0.999588,1.508427)
  (1.022053,1.476938)
  (1.048521,1.450868)
  (1.075136,1.424387)
  (1.101942,1.397483)
  (1.128985,1.370142)
  (1.156311,1.342354)
  (1.184402,1.315012)
  (1.212417,1.286317)
  (1.241255,1.258074)
  (1.270529,1.229379)
  (1.299925,1.199300)
  (1.330234,1.169695)
  (1.360801,1.138697)
  (1.392036,1.107244)
  (1.423989,1.075346)
  (1.456710,1.043013)
  (1.489999,1.009290)
  (1.524194,0.975152)
  (1.559343,0.940617)
  (1.595117,0.903743)
  (1.632191,0.867487)
  (1.670105,0.828921)
  (1.709286,0.790039)
  (1.749670,0.749882)
  (1.791365,0.708478)
  (1.834404,0.664864)
  (1.879004,0.620074)
  (1.925269,0.574151)
  (1.973290,0.526144)
  (2.022806,0.522105)
  (2.070984,0.570037)
  (2.117353,0.616888)
  (2.162084,0.661611)
  (2.205253,0.705161)
  (2.247073,0.746504)
  (2.287581,0.786604)
  (2.326734,0.826420)
  (2.364917,0.863947)
  (2.401920,0.901136)
  (2.437999,0.936983)
  (2.473035,0.972449)
  (2.507323,1.006543)
  (2.540705,1.040223)
  (2.573515,1.072514)
  (2.605557,1.104373)
  (2.636880,1.135789)
  (2.667534,1.166752)
  (2.697567,1.197255)
  (2.727409,1.226370)
  (2.756368,1.255950)
  (2.785269,1.284156)
  (2.813347,1.312814)
  (2.841048,1.341013)
  (2.868888,1.367874)
  (2.895992,1.395182)
  (2.922859,1.422053)
  (2.949533,1.448501)
  (2.975520,1.475382)
  (3.000258,1.503510)
  (3.000231,1.566458)
  (3.000452,1.625556)
  (3.000534,1.681684)
  (3.000023,1.735683)
  (3.000307,1.786033)
  (3.000289,1.834359)
  (3.000080,1.880696)
  (3.000471,1.924352)
  (3.000220,1.966848)
  (2.000000,3.000000)
	
	\rput(2.0,2.0){\diamond}

	\psset{linecolor=black!60!white}
	\pspolygon*
	(1.499765,2.484938)
  (1.499932,2.464070)
  (1.499403,2.441518)
  (1.499580,2.418735)
  (1.499836,2.394970)
  (1.499604,2.369422)
  (1.499593,2.342833)
  (1.499860,2.315191)
  (1.499891,2.285665)
  (1.499794,2.254213)
  (1.499680,2.220798)
  (1.499662,2.185385)
  (1.499854,2.147945)
  (1.499884,2.107580)
  (1.499921,2.064243)
  (1.499683,2.016998)
  (1.510729,1.989187)
  (1.527283,1.971427)
  (1.544753,1.954844)
  (1.562317,1.937628)
  (1.580027,1.919767)
  (1.597934,1.901254)
  (1.616416,1.883029)
  (1.635160,1.864156)
  (1.654221,1.844633)
  (1.673919,1.825423)
  (1.694000,1.805574)
  (1.714516,1.785091)
  (1.735522,1.763979)
  (1.757070,1.742246)
  (1.779213,1.719901)
  (1.802003,1.696957)
  (1.825493,1.673429)
  (1.849736,1.649333)
  (1.874782,1.624689)
  (1.900614,1.598520)
  (1.927392,1.571844)
  (1.955168,1.544690)
  (1.983968,1.515087)
  (2.013686,1.513063)
  (2.042602,1.541622)
  (2.070431,1.569733)
  (2.097292,1.596368)
  (2.123205,1.622499)
  (2.148222,1.648101)
  (2.172523,1.672161)
  (2.196070,1.695654)
  (2.218917,1.718565)
  (2.241115,1.740878)
  (2.262716,1.762581)
  (2.283775,1.783664)
  (2.304344,1.804119)
  (2.324476,1.823941)
  (2.344223,1.843125)
  (2.363334,1.862624)
  (2.382128,1.881473)
  (2.400657,1.899675)
  (2.418613,1.918167)
  (2.436370,1.936007)
  (2.453980,1.953203)
  (2.471078,1.970675)
  (2.488095,1.987508)
  (2.500107,2.013527)
  (2.500468,2.059917)
  (2.500157,2.104166)
  (2.500302,2.144560)
  (2.500084,2.182883)
  (2.500156,2.218317)
  (2.500129,2.251755)
  (2.500116,2.283229)
  (2.500226,2.312778)
  (2.500569,2.340444)
  (2.500011,2.367841)
  (2.500471,2.392631)
  (2.500144,2.417180)
  (2.500373,2.439975)
  (2.500576,2.461810)
  (2.500110,2.483424)
  (2.000000,3.000000)

	\psset{linestyle=solid,linecolor=black!80!white,arrowsize=4pt,arrowinset=0.1,arrowlength=1.0}
	
  \psline[linewidth=0.7pt]{->}(0.0,2.0)(3.96,2.0)
  \psline[linewidth=0.7pt]{->}(2.0,0.1)(2.0,4.0)
		
	\rput(2.0,2.5){\pscircle*[linecolor=black!5!white](0,0){0.067}}
	\rput(2.0,1.5){\pscircle*[linecolor=black!5!white](0,0){0.067}}	
			
	\rput(2.0,3.0){\rnode{x}{\pscircle*(0,0){0.067}}}
	
	
	\psset{linestyle=solid,linecolor=black,linewidth=0.4pt}
	
	\rput[l](2.16,3.05){$x^\star$}
	
	\rput[l](2.82,2.57){$w^{(c)}$}
	\psline{-}(2.1,2.5)(2.76,2.5)
	
	\rput[r](0.8,1.57){$w^{(d)}$}
	\psline{-}(0.82,1.5)(1.9,1.5)
		
	\rput[rb](1.83,2.72){\small $S_{f_1}^{(c)}$}	
	\rput[rt](1.8,0.75){\small $S_{f_1}^{(d)}$}	
	\rput[l](3.1,1.16){\small $S_{\|\cdot\|_1}$}
	\psline{-}(2.22,1.2)(3.05,1.2)
	
	\end{pspicture}
	}
	}
	\subfigure[$\ell_1$-$\ell_2$: bad components]{\label{SubFig:IllustrationL1L2Bad}
	\psscalebox{1.0}{
	\begin{pspicture}(3.96,3.9)
		\psset{blendmode=2}	
	
	\def\radius{1}       
								
	\def\diamond{       
		\pspolygon*[linecolor=black!35!white](0,\radius)(\radius,0.0)(0.0,-\radius)(-\radius,0.0)	
	}

	\psset{linecolor=black!20!white}
	\pspolygon*
	(1.903289,2.959111)
  (1.805512,2.913147)
  (1.709575,2.863273)
  (1.616490,2.809990)
  (1.525458,2.752948)
  (1.436592,2.692209)
  (1.350003,2.627837)
  (1.265801,2.559900)
  (1.184941,2.488999)
  (1.106658,2.414710)
  (1.031881,2.337674)
  (0.959863,2.257437)
  (0.891504,2.174676)
  (0.826072,2.088908)
  (0.764437,2.000848)
  (0.771663,1.965386)
  (0.780571,1.930478)
  (0.789650,1.894842)
  (0.799672,1.859146)
  (0.810642,1.823414)
  (0.821869,1.786951)
  (0.834084,1.750472)
  (0.847292,1.714001)
  (0.860844,1.676807)
  (0.875428,1.639646)
  (0.891048,1.602545)
  (0.907706,1.565530)
  (0.924816,1.527822)
  (0.943003,1.490230)
  (0.961710,1.451953)
  (0.981530,1.413827)
  (1.002464,1.375882)
  (1.024514,1.338147)
  (1.047677,1.300652)
  (1.071483,1.262543)
  (1.095983,1.223824)
  (1.122100,1.186298)
  (1.148942,1.148205)
  (1.177361,1.111386)
  (1.206153,1.073121)
  (1.236512,1.036194)
  (1.268035,0.999718)
  (1.300389,0.962781)
  (1.333931,0.926347)
  (1.368653,0.890449)
  (1.404548,0.855119)
  (1.441357,0.819422)
  (1.479587,0.785323)
  (1.518751,0.750911)
  (1.559275,0.718172)
  (1.600747,0.685179)
  (1.643365,0.652941)
  (1.687114,0.621491)
  (1.731979,0.590863)
  (1.777945,0.561088)
  (1.824996,0.532197)
  (1.873115,0.504223)
  (1.922282,0.477197)
  (1.972481,0.451149)
  (2.023478,0.449108)
  (2.073758,0.475077)
  (2.123009,0.502027)
  (2.171212,0.529927)
  (2.218348,0.558745)
  (2.264402,0.588451)
  (2.309356,0.619013)
  (2.353194,0.650398)
  (2.395902,0.682574)
  (2.437467,0.715509)
  (2.478082,0.748191)
  (2.517339,0.782549)
  (2.555663,0.816597)
  (2.592568,0.852246)
  (2.628557,0.887529)
  (2.663374,0.923384)
  (2.697012,0.959777)
  (2.729463,0.996676)
  (2.761081,1.033115)
  (2.791157,1.070934)
  (2.820426,1.108243)
  (2.848523,1.145943)
  (2.875879,1.183100)
  (2.901639,1.221494)
  (2.926688,1.259301)
  (2.950590,1.297391)
  (2.973848,1.334869)
  (2.995470,1.373443)
  (3.016481,1.411365)
  (3.036379,1.449470)
  (3.055736,1.486908)
  (3.074015,1.524494)
  (3.091217,1.562199)
  (3.107345,1.599996)
  (3.123040,1.637085)
  (3.137698,1.674235)
  (3.151325,1.711421)
  (3.164606,1.747884)
  (3.176893,1.784358)
  (3.188903,1.820112)
  (3.199234,1.856546)
  (3.209326,1.892240)
  (3.219225,1.927216)
  (3.227451,1.962786)
  (3.236300,1.996991)
  (3.178788,2.082125)
  (3.114391,2.167525)
  (3.045456,2.251123)
  (2.973658,2.331606)
  (2.899094,2.408894)
  (2.821016,2.483441)
  (2.740352,2.554607)
  (2.657206,2.622318)
  (2.570808,2.686984)
  (2.482125,2.748024)
  (2.391265,2.805371)
  (2.297440,2.859392)
  (2.202561,2.909154)
  (2.104930,2.955439)
  (2.000000,3.000000)
	
	\rput(2.0,2.0){\diamond}

	\psset{linecolor=black!60!white}
	\pspolygon*
	(1.905590,2.932600)
  (1.849206,2.887727)
  (1.794602,2.840590)
  (1.741846,2.791241)
  (1.691769,2.740380)
  (1.642892,2.686792)
  (1.596796,2.631839)
  (1.552051,2.574247)
  (1.510181,2.515440)
  (1.471200,2.455527)
  (1.434439,2.393880)
  (1.399283,2.329821)
  (1.367788,2.265651)
  (1.338637,2.199973)
  (1.311878,2.132857)
  (1.287555,2.064372)
  (1.268663,1.998628)
  (1.285394,1.979287)
  (1.302608,1.960217)
  (1.320300,1.941430)
  (1.337942,1.922083)
  (1.356088,1.903022)
  (1.374733,1.884257)
  (1.393872,1.865802)
  (1.413047,1.846778)
  (1.432741,1.828069)
  (1.452951,1.809690)
  (1.473271,1.790737)
  (1.494132,1.772123)
  (1.515528,1.753864)
  (1.537453,1.735972)
  (1.559576,1.717517)
  (1.582254,1.699445)
  (1.605480,1.681770)
  (1.628979,1.663545)
  (1.653051,1.645736)
  (1.677687,1.628360)
  (1.702672,1.610454)
  (1.728245,1.593004)
  (1.754397,1.576025)
  (1.781121,1.559534)
  (1.808277,1.542556)
  (1.836025,1.526093)
  (1.864265,1.509166)
  (1.893116,1.492785)
  (1.922569,1.476967)
  (1.952612,1.461730)
  (1.983223,1.446091)
  (2.014312,1.445066)
  (2.044967,1.460657)
  (2.075055,1.475847)
  (2.104552,1.491619)
  (2.133450,1.507956)
  (2.161736,1.524840)
  (2.189530,1.541261)
  (2.216733,1.558199)
  (2.243503,1.574650)
  (2.269702,1.591591)
  (2.295322,1.609004)
  (2.320355,1.626875)
  (2.345039,1.644216)
  (2.369158,1.661992)
  (2.392705,1.680185)
  (2.415978,1.697828)
  (2.438704,1.715871)
  (2.460876,1.734298)
  (2.482848,1.752161)
  (2.503912,1.771319)
  (2.525200,1.788982)
  (2.545569,1.807911)
  (2.565826,1.826266)
  (2.585568,1.844952)
  (2.604791,1.863956)
  (2.623977,1.882390)
  (2.642164,1.901998)
  (2.660862,1.920177)
  (2.678552,1.939506)
  (2.696291,1.958277)
  (2.712979,1.978150)
  (2.730330,1.996656)
  (2.713976,2.059256)
  (2.690462,2.127056)
  (2.663903,2.194290)
  (2.634949,2.260091)
  (2.602980,2.325137)
  (2.568682,2.388587)
  (2.532106,2.450374)
  (2.492596,2.511139)
  (2.450889,2.570083)
  (2.407041,2.627144)
  (2.361113,2.682260)
  (2.312400,2.736017)
  (2.261704,2.787680)
  (2.209878,2.836576)
  (2.155422,2.883896)
  (2.099180,2.928956)
  (2.000000,3.000000)
		
	\psset{linestyle=solid,linecolor=black!80!white,arrowsize=4pt,arrowinset=0.1,arrowlength=1.0}
	
  \psline[linewidth=0.7pt]{->}(0.0,2.0)(3.96,2.0)
  \psline[linewidth=0.7pt]{->}(2.0,0.1)(2.0,4.0)
		
	\rput(2.0,2.5){\pscircle*[linecolor=black!5!white](0,0){0.067}}
	\rput(2.0,1.5){\pscircle*[linecolor=black!5!white](0,0){0.067}}	
			
	\rput(2.0,3.0){\rnode{x}{\pscircle*(0,0){0.067}}}
	
	
	\psset{linestyle=solid,linecolor=black,linewidth=0.4pt}
	
	\rput[l](2.16,3.05){$x^\star$}
	
	\rput[l](3.1,2.57){$w^{(c)}$}
	\psline{-}(2.1,2.5)(3.05,2.5)
	
	\rput[r](0.8,1.57){$w^{(d)}$}
	\psline{-}(0.82,1.5)(1.9,1.5)
		
	\rput[r](0.8,2.7){\small $S_{f_2}^{(c)}$}	
	\psline{-}(0.82,2.7)(1.72,2.7)
	\rput[rt](1.8,0.57){\small $S_{f_2}^{(d)}$}	
	\rput[l](3.1,1.16){\small $S_{\|\cdot\|_1}$}
	\psline{-}(2.22,1.2)(3.05,1.2)
	
	\end{pspicture}
	}
	}	
	\caption{
		Sublevel sets $S_{f_i}^{(j)} := \{x\,:\, \|x\|_1 + g_i(x-w^{(j)}) \leq \|x^\star\|_1 + g_i(x^\star-w^{(j)})\}$, for~$i=1,2$ and $j = a,b,c,d$, where $w^{(a)} = (0,1.6)$, $w^{(b)} = (0,1.3)$, $w^{(c)} = (0,0.5)$, and $w^{(d)} = (0,-0.5)$. In both \text{(a)} and \text{(b)}, $w$ is $w^{(a)}$ and~$w^{(b)}$, whose nonzero components are good. In both \text{(c)} and \text{(d)}, $w$ is $w^{(c)}$ and~$w^{(d)}$, whose nonzero components are bad. The $\ell_1$-norm sublevel set~$S_{\|\cdot\|_1}$ at~$x^\star$ is also shown in all figures, and is associated with~\eqref{Eq:BP}.
	}
	\label{Fig:GoodAndBadComponents}
	
	\end{figure*}

\mypar{Good and bad components}	
	Naturally, our bounds for $\ell_1$-$\ell_1$ and $\ell_1$-$\ell_2$ minimization are a function of the ``quality'' of the side information~$w$. A way to measure the quality of each component of~$w$ arises naturally in the proofs of our results, but it can be motivated geometrically, as we do next. First, recall that the relation~$T_f(x^\star) = \text{cone}(S_f(x^\star) - x^\star)$ means that~$T_f(x^\star)$ is composed of all the half-lines that join~$x^\star$ to a point of the sublevel set~$S_f(x^\star)$.	
	Therefore, the width of~$T_f(x^\star)$ can be estimated by looking at the sublevel set~$S_f(x^\star)$. \fref{Fig:GoodAndBadComponents} shows the sublevel sets of~$f_1$ and~$f_2$ for~$n = 2$. In those plots, $x^\star$ is always $x^\star = (0,1)$, and we consider four different $w$'s: $w^{(a)} = (0,1.6)$, $w^{(b)} = (0,1.3)$, $w^{(c)} = (0,0.5)$, and $w^{(d)} = (0,-0.5)$. In Figs.~\ref{SubFig:IllustrationL1L1Good} and~\ref{SubFig:IllustrationL1L2Good} the side information is~$w^{(a)}$ and~$w^{(b)}$, and in Figs.~\ref{SubFig:IllustrationL1L1Bad} and~\ref{SubFig:IllustrationL1L2Bad} it is~$w^{(c)}$ and~$w^{(d)}$. Figs.~\ref{SubFig:IllustrationL1L1Good} and~\ref{SubFig:IllustrationL1L1Bad} show the sublevel sets of~$f_1$, while Figs.~\ref{SubFig:IllustrationL1L2Good} and~\ref{SubFig:IllustrationL1L2Bad} show the sublevel sets of~$f_2$. For reference, we show in all plots the sublevel set of the $\ell_1$-norm ball $S_{\|\cdot\|_1}:= B_2(0,\|x^\star\|_1)$. To represent all the other sublevel sets, we use the notation $S_{f_i}^{(j)} := \{x:\, \|x\|_1 + g_i(x-w^{(j)}) \leq \|x^\star\|_1 + g_i(x^\star-w^{(j)})\}$, for~$i=1,2$ and $j = a,b,c,d$.	For example, the sublevel sets in \fref{SubFig:IllustrationL1L1Good} are the line segments $S_{f_1}^{(a)} = \{(0,x_2):\, 0 \leq x_2 \leq 1.6\}$ and $S_{f_1}^{(b)} = \{(0,x_2):\, 0 \leq x_2 \leq 1.3\}$. The tangent cone they generate is the line $\{(0,x_2):\, x_2 \in \mathbb{R}\}$, which has zero Gaussian width. This means that the nonzero components of~$w^{(a)}$ and~$w^{(b)}$ do not contribute ``any width'' to $T_{f_1}(x^\star)$.	A careful inspection of the remaining figures reveals that the tangent cones in Figs.~\ref{SubFig:IllustrationL1L1Good} and~\ref{SubFig:IllustrationL1L2Good} have smaller ``geometrical widths'' (and thus Gaussian widths) than the cone generated by~$S_{\|\cdot\|_1}$. In Figs.~\ref{SubFig:IllustrationL1L1Bad} and~\ref{SubFig:IllustrationL1L2Bad}, in contrast, the tangent cones have either the same width as the cone generated by~$S_{\|\cdot\|_1}$ (\fref{SubFig:IllustrationL1L1Bad}), or larger widths (\fref{SubFig:IllustrationL1L2Bad}). Note, in particular that, in \fref{SubFig:IllustrationL1L1Bad}, $S_{f_1}^{(c)}$, $S_{f_1}^{(d)}$, and~$S_{\|\cdot\|_1}$ all generate the same tangent cone.
	In \fref{SubFig:IllustrationL1L2Bad}, $S_{f_2}^{(c)}$ and~$S_{f_2}^{(d)}$ generate tangent cones with widths larger than the cone generated by~$S_{\|\cdot\|_1}$. Since we want small widths, we say that~$w_2$, the nonzero component of~$w$, is a \textit{good component} in Figs.~\ref{SubFig:IllustrationL1L1Good} and~\ref{SubFig:IllustrationL1L2Good} and is a \textit{bad component} in Figs.~\ref{SubFig:IllustrationL1L1Bad} and~\ref{SubFig:IllustrationL1L2Bad}. The generic definition is:
	\begin{Definition}[Good and bad components]
	\label{def:GoodBadComponents}
		Let~$x^\star \in \mathbb{R}^n$ be the vector to reconstruct and let~$w \in \mathbb{R}^n$ be the side information. For $i=1,\ldots,n$, a component~$w_i$ is considered good if		
		$$
			\text{$x_i^\star > 0$\, \text{and} \, $x_i^\star < w_i$}
			\qquad
			\text{or}
			\qquad
			\text{$x_i^\star < 0$\, \text{and} \, $x_i^\star > w_i$}\,,
		$$		
		and~$w_i$ is considered bad if		
		$$
			\text{$x_i^\star > 0$\, \text{and} \, $x_i^\star > w_i$}
			\qquad
			\text{or}
			\qquad
			\text{$x_i^\star < 0$\, \text{and} \, $x_i^\star < w_i$}\,.
		$$
	\end{Definition}		
	\fref{Fig:GoodAndBadComponents} gives the intuition why $\ell_1$-$\ell_1$ minimization requires less measurements than standard CS and $\ell_1$-$\ell_2$ minimization: its good components induce less width in~$T_{f_1}(x^\star)$, and its bad components never induce more width in~$T_{f_1}(x^\star)$ than the absence of side information. 
	Although \fref{Fig:GoodAndBadComponents} shows the impact only of the components~$w_i$ for which~$x_i^\star \neq 0$ and~$x_i^\star \neq w_i$, the components for which $w_i = x_i^\star \neq 0$ and for which~$w_i \neq x_i^\star = 0$ also impact the Gaussian width, as shown next.

\section{Bounds For $\ell_1$-$\ell_1$ and $\ell_1$-$\ell_2$ Minimization}

	To state our results for $\ell_1$-$\ell_1$ minimization, we need to define
	\begin{align*}
		\overline{h} 
			&:= 
			\big|
				\{i\,:\, x_i^\star > 0, \,\,x_i^\star > w_i\} \cup \{i\,:\, x_i^\star < 0, \,\,x_i^\star < w_i\}
			\big|
		\\
		\xi &:= \big|\{i\,:\, w_i \neq x_i^\star = 0\}\big| - \big|\{i\,:\, w_i = x_i^\star \neq 0\}\big|\,,	
	\end{align*}
	where~$|\cdot|$ denotes the cardinality of a set. Note that~$\overline{h}$ is the number of bad components of~$w$. Naturally, $\overline{h} \leq s$, where the difference~$s - \overline{h} = h + r$ is the number of good components~$h$ plus $r:=|\{i\,:\, w_i = x_i^\star \neq 0\}|$. The quantity~$\xi$ is the number of components where~$w$ overestimates the support of~$x^\star$ minus~$r$.	Our bound for $\ell_1$-$\ell_1$ minimization depends on these two key parameters.
	
	\begin{Theorem}[$\ell_1$-$\ell_1$ minimization]	
	\label{Thm:L1L1Reconstruction}
		Let~$x^\star \in \mathbb{R}^n$ be the vector to reconstruct and let~$w \in \mathbb{R}^n$ be the side information. Let~$f_1(x) = \|x\|_1 + \|x - w\|_1$, and assume~$\overline{h} > 0$ and that there exists at least one index~$i$ for which~$x_i^\star = w_i = 0$. Let the entries of~$A \in \mathbb{R}^{m \times n}$ be i.i.d.\ Gaussian with zero mean and variance~$1/m$. Then, 
		\begin{equation}\label{Eq:ThmL1L1}
			w\bigl(T_{f_1}(x^\star)\bigr)^2 
			\leq 
			2\overline{h}\log\Big(\frac{n}{s + \xi/2}\Big) + \frac{7}{5}\Big(s + \frac{\xi}{2}\Big)\,.
		\end{equation}
		Namely, if~$m \geq 2\overline{h}\log\big(n/(s + \xi/2) + (7/5)(s + \xi/2) + 1$, then $x^\star$ is the unique solution of~\eqref{Eq:BPSideInfoGeneric} with~$g = g_1$ and~$\beta = 1$, with probability at least $1 - \exp\big(-\frac{1}{2}(\lambda_m - m)^2\big)$.
	\end{Theorem}

	By upper bounding the squared Gaussian width of the tangent cone of~$f_1$, \thref{Thm:L1L1Reconstruction} provides a number of measurements above which $\ell_1$-$\ell_1$ minimization reconstructs~$x^\star$ with high probability. The assumption that there is at least one bad component, $\overline{h}>0$, is necessary to guarantee that the subdifferential~$\partial f_1(x^\star)$ equals the normal cone of~$f_1$ at~$x^\star$~\cite{Lemarechal04-FundamentalsConvexAnalysis}, a relation used in the proof. When~$\beta\neq 1$, the assumption~$\overline{h} > 0$ can be relaxed; see~\cite{Mota14-CompressedSensingSideInformation}. It can be shown that if, contrary to the theorem's assumptions, there is no index~$i$ for which~$x_i^\star = w_i = 0$, we can have~$n = s + \xi/2$, making the right-hand side of~\eqref{Eq:ThmL1L1} evaluate to~$-\infty$~\cite{Mota14-CompressedSensingSideInformation}. 
		
	Notice that~\eqref{Eq:PropChandrasekaranBound} and~\eqref{Eq:ThmL1L1} have the same format and both provide reconstruction guarantees with probability at least $1 - \exp\big(-\frac{1}{2}(\lambda_m - m)^2\big)$. To compare~\eqref{Eq:PropChandrasekaranBound} and~\eqref{Eq:ThmL1L1}, assume first that~$\xi = 0$. In that case, both bounds are equal, apart from the terms multiplying the $\log$'s: $2s$ in~\eqref{Eq:PropChandrasekaranBound} and~$2\overline{h}$ in~\eqref{Eq:ThmL1L1}. Since~$s - \overline{h} = h + r\geq 0$, the larger the number of good components, $h$, and the larger~$r$ (number of components where~$x^\star$ and~$w$ coincide on the support of~$x^\star$), the smaller is~\eqref{Eq:ThmL1L1} with respect to~\eqref{Eq:PropChandrasekaranBound}. This confirms the interpretation given in \fref{Fig:GoodAndBadComponents} and complements it with the intuitive fact that a large~$r$ should decrease the number of measurements.\footnote{Given that~$s - \overline{h} = h + r$, we could have defined the good components as the components~$i$ for which $x_i^\star > 0$ and~$x_i^\star \leq w_i$, or $x_i^\star$ and~$x_i^\star \geq w_i$. In that case, $s-\overline{h}$ would be exactly the number of good components. This was not done in~\cite{Mota14-CompressedSensingSideInformation} for technical reasons, and we kept the same notation here.} In general, however, $\xi \neq 0$. In that case, if~$n$ is much larger than~$\xi$ and~$s$, if~$h,r> 0$, and if~$\xi$ is larger than a small negative number, then~$2\overline{h}\log(n/(s + \xi/2))$, the dominant term of~\eqref{Eq:ThmL1L1}, is smaller than $2s\log(n/s)$, the dominant term of~\eqref{Eq:PropChandrasekaranBound}. That is, \eqref{Eq:ThmL1L1} is asymptotically smaller than~\eqref{Eq:PropChandrasekaranBound}. 
						
	To present our results on $\ell_1$-$\ell_2$ minimization, we define 
	\begin{align*}
				I &:= \bigl\{i\,:\, x_i^\star \neq 0\bigr\}
			&
				J &:= \bigl\{j\,:\, x_j^\star \neq w_j\bigr\}
			\\
				I_+ &:= \bigl\{i\,:\, x_i^\star > 0\bigr\}
			&
				I_- &:= \bigl\{i\,:\, x_i^\star < 0\bigr\}\,,
			\,,
	\end{align*}
	and~$K := |\{i \in I^c \cap J\,:\, |w_i| \geq 1\}|$, where~$I^c$ is the complement of~$I$.
	We also define~$q:= |I \cup J|$ and~$\overline{w}:=\max_{i \in I^c}|w_i|$. 
	Note that~$\overline{w} \leq \|w\|_\infty$.
	\begin{Theorem}[$\ell_1$-$\ell_2$ minimization]
	\label{Thm:L1L2Reconstruction}
		Let~$x^\star, w \in \mathbb{R}^n$ be as in \thref{Thm:L1L1Reconstruction}. Let~$f_2(x) = \|x\|_1 + \frac{1}{2}\|x - w\|_2^2$ and assume~$x^\star\neq 0$, $q < n$, and that either~$\overline{w} < 1$ or that there exists~$i\in I \cap J$ such that $\beta \neq \text{\emph{sign}}(x_i^\star)/(w_i - x_i^\star)$. Assume also that
		\begin{equation}\label{Eq:ThmL1L2Assumption}
			\frac{q-s}{n-q} \leq |1 - \overline{w}|\exp\Big(2\overline{w}\log\Big(\frac{n}{q}\Big)\Big(\frac{\overline{w}}{2}-1\Big)\Big)\,.
		\end{equation} 
		Then,
		\begin{equation}\label{Eq:ThmL1L2Bound}
			  w\bigl(T_{f_2}(x^\star)\bigr)^2
			\leq			
			  2 v \log\Big(\frac{n}{q}\Big) + s + 2K + \frac{4}{5}q\,,
		\end{equation}
		where
		$$
			v 
			:=
			\sum_{i \in I_+} (1 + x_i^\star - w_i)^2 
			+
			\sum_{i \in I_-} (1 + w_i - x_i^\star)^2 
			+
			\sum_{i \in I\cap J^c} (|w_i| - 1)^2\,.			
		$$
	\end{Theorem}
	\thref{Thm:L1L2Reconstruction} not only requires assumptions stronger than the ones in \thref{Thm:L1L1Reconstruction}, but also provides a larger bound. The assumption~$q < n$ makes the right-hand sides of~\eqref{Eq:ThmL1L2Assumption} and~\eqref{Eq:ThmL1L2Bound} finite. The assumption that~$\overline{w} < 1$ or that there exists~$i \in I \cap J$ such that $\beta \neq \text{sign}(x_i^\star)/(w_i - x_i^\star)$ guarantees that~$\partial f_2(x^\star)$ equals the normal cone of~$f_2$ at~$x^\star$~\cite{Lemarechal04-FundamentalsConvexAnalysis}. The case where assumption~\eqref{Eq:ThmL1L2Assumption} does not hold is also addressed in~\cite{Mota14-CompressedSensingSideInformation}. Note, however, that it is easy to satisfy~\eqref{Eq:ThmL1L2Assumption} in practice: its left-hand side can be shown to equal~$|I^c \cap J|/|I^c \cap J^c|$, i.e., the number of components in which~$x^\star$ and~$w$ differ outside~$I$ divided by the number of components in which~$x^\star$ and~$w$ are both zero; if~$x^\star$ and~$w$ are sparse enough, this number is smaller than~$1$. And the right-hand side of~\eqref{Eq:ThmL1L2Assumption} can be large whenever~$\overline{w} \neq 1$. The bound in~\eqref{Eq:ThmL1L2Bound} has the same format as~\eqref{Eq:PropChandrasekaranBound} and~\eqref{Eq:ThmL1L1}. The parameter~$v$ in~\eqref{Eq:ThmL1L2Bound}, however, depends on the magnitude of the components of both~$x^\star$ and~$w$. This is contrast with~\eqref{Eq:PropChandrasekaranBound} and~\eqref{Eq:ThmL1L1}, whose parameters depend only on the signs of~$x^\star$ and~$x^\star - w$, but not on the magnitudes of their components. This was expected from the interpretation of \fref{Fig:GoodAndBadComponents}: the widths of the tangent cones in Figs.~\ref{SubFig:IllustrationL1L1Good} and~\ref{SubFig:IllustrationL1L1Bad} do not vary with the magnitude of the nonzero component of~$w$, whereas the widths in Figs.~\ref{SubFig:IllustrationL1L2Good} and~\ref{SubFig:IllustrationL1L2Bad} do. It is not clear when~\eqref{Eq:ThmL1L2Bound} is smaller than~\eqref{Eq:PropChandrasekaranBound}, but note that the term inside $\log(\cdot)$ is smaller in~\eqref{Eq:ThmL1L2Bound}, and~$v$ is always larger than~$s$ (e.g., if~$w_i = x_i^\star$ for all $i \in I$,  $v = s + \sum_{I\cap J^c}(|w_i| - 1)^2 \geq s$). 
	
\begin{figure}[t]
	\centering
		
	\readdata{\data}{figures/ResultsLaTexSI.dat}
	\readdata{\dataCS}{figures/ResultsLaTexCS.dat}
	\readdata{\dataSILT}{figures/ResultsLaTexSIL2.dat}
		
	\psscalebox{1.0}{
	\begin{pspicture}(8.8,4.95)
					
		\def\xMax{700}                                
		\def\xMin{0}                                  
		\def\xNumTicks{7}                             
		\def\yMax{1.00}                               
		\def\yMin{0}                                  
		\def\yNumTicks{5}                             
		\def\xIncrement{100}                          
		\def\yIncrement{0.2}                          
					
		\def\xOrig{0.50}                              
		\def\yOrig{0.80}                              
		\def\SizeX{8.00}                              
		\def\SizeY{3.70}                              
		\def\xTickIncr{1.14}                          
		\def\yTickIncr{0.74}                          

		\input{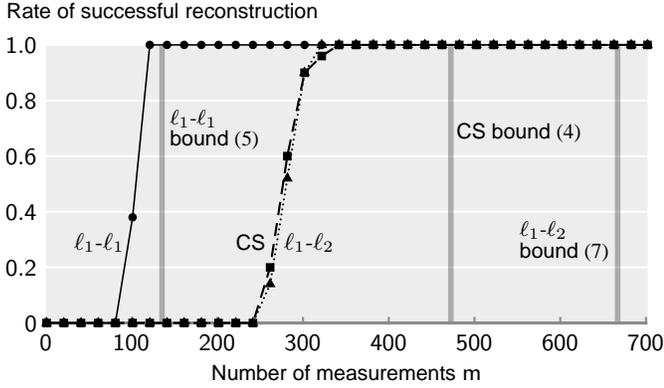}
		
		
		\psset{xunit=\xScale\psunit,yunit=\yScale\psunit,linewidth=0.8pt}
		\psline[linewidth=2.1pt,linecolor=black!90!white,strokeopacity=0.31]{-}(179,0.222)(179,1.212)		
		\psline[linewidth=2.1pt,linecolor=black!90!white,strokeopacity=0.31]{-}(515,0.222)(515,1.212)
		\psline[linewidth=2.1pt,linecolor=black!90!white,strokeopacity=0.31]{-}(709,0.222)(709,1.212)
		\dataplot[origin={\xDataOrig,\yDataOrig},showpoints=true,linestyle=dotted,dotsep=1.1pt,dotstyle=triangle*]{\dataSILT}
		\dataplot[origin={\xDataOrig,\yDataOrig},showpoints=true,linestyle=dashed,fillcolor=black!40!white,dotstyle=square*]{\dataCS}			
		\dataplot[origin={\xDataOrig,\yDataOrig},showpoints=true,linewidth=0.6pt]{\data}
		\psset{xunit=\psunit,yunit=\psunit}
		
		\rput[lb](-0.02,4.85){\footnotesize \textbf{\sf Rate of successful reconstruction}}
		\rput[ct](4.5,0.14){\footnotesize \textbf{\sf Number of measurements {\small $\mathsf{m}$}}}
			
		\rput[rt](1.54,2.0){\small $\ell_1$-$\ell_1$}
		\rput[lt](3.67,2.00){\small $\ell_1$-$\ell_2$}
		\rput[rt](3.42,1.98){\footnotesize {\sf CS}}
		\rput[lt](2.15,3.64){\footnotesize $\ell_1$-$\ell_1$}
		\rput[lt](2.15,3.36){\sf\footnotesize bound \eqref{Eq:ThmL1L1}}
		\rput[lb](5.96,3.2){\footnotesize {\sf CS bound \eqref{Eq:PropChandrasekaranBound}}}
		\rput[lt](6.80,2.15){\footnotesize $\ell_1$-$\ell_2$ }	
		\rput[lt](6.80,1.87){\sf\footnotesize bound \eqref{Eq:ThmL1L2Bound}}
	\end{pspicture}
	}
	\vspace{-0.6cm}
	\caption{
		Reconstruction rate of standard CS~\eqref{Eq:BP}, $\ell_1$-$\ell_1$ minimization, and $\ell_1$-$\ell_2$ minimization. The vertical lines are the bounds in~\eqref{Eq:PropChandrasekaranBound}, \eqref{Eq:ThmL1L1}, and~\eqref{Eq:ThmL1L2Bound}. 
	}
	\label{Fig:ExperimentalResults}
\end{figure}

\section{Experimental Results}

\begin{figure}[t]
	\centering
			
	\readdata{\betaA}{figures/BetaResults1.dat}
	\readdata{\betaB}{figures/BetaResults2.dat}
	\readdata{\betaC}{figures/BetaResults3.dat}
	\readdata{\betaD}{figures/BetaResults4.dat}
	\readdata{\betaE}{figures/BetaResults5.dat}
	\readdata{\bound}{figures/boundsL1L1.dat}
	
	\psscalebox{1.0}{
	\begin{pspicture}(8.8,4.95)	
					
		\def\xMax{2}                                
		\def\xMin{-2}                               
		\def\xNumTicks{4}                           
		\def\yMax{350}                              
		\def\yMin{0}                                
		\def\yNumTicks{5}                           
		\def\xIncrement{1}                          
		\def\yIncrement{70}                         
					
		\def\xOrig{0.50}                              
		\def\yOrig{0.80}                              
		\def\SizeX{8.00}                              
		\def\SizeY{3.70}                              
		\def\xTickIncr{2.00}                          
		\def\yTickIncr{0.74}                          

		\definecolor{colorXAxis}{gray}{0.55}           
		\definecolor{colorBackground}{gray}{0.93}      

		\def \distXLabels{0.15}

		\def \distYLabels{0.12}

		\def \xTickWidth{0.08}

		\fpAdd{\xNumTicks}{1}{\xNumTicksPOne}         
		\fpAdd{\yNumTicks}{1}{\yNumTicksPOne}

		\FPadd \xEndPoint \xOrig \SizeX                
		\FPadd \yEndPoint \yOrig \SizeY                

		\FPset \unit 1

		\FPsub \xRange \xMax \xMin			
		\FPdiv \xScale \SizeX \xRange 
		\FPdiv \xMultByOrigin \unit \xScale
		\FPmul \xDataOrig  \xMultByOrigin \xOrig

		\FPsub \yRange \yMax \yMin
		\FPdiv \yScale \SizeY \yRange 
		\FPdiv \yMultByOrigin \unit \yScale
		\FPmul \yDataOrig  \yMultByOrigin \yOrig

		\fpSub{\yOrig}{\distXLabels}{\xPosLabels}
		\fpSub{\xOrig}{\distYLabels}{\yPosLabels}

		\fpAdd{\yOrig}{\xTickWidth}{\xTickTop}

		\psframe*[linecolor=colorBackground](\xOrig,\yOrig)(\xEndPoint,\yEndPoint)
									
		\multido{\nx=\xOrig+\xTickIncr, \iB=\xMin+\xIncrement}{\xNumTicksPOne}{	
			\rput[t](\nx,\xPosLabels){\small $\mathsf{10^{\iB}}$}
			\psline[linewidth=0.8pt,linecolor=colorXAxis](\nx,\yOrig)(\nx,\xTickTop)	
		}						

		\multido{\ny=\yOrig+\yTickIncr, \nB=\yMin+\yIncrement}{\yNumTicksPOne}{	
			\rput[r](\yPosLabels,\ny){\small $\mathsf{\nB}$}
			\psline[linecolor=white,linewidth=0.8pt](\xOrig,\ny)(\xEndPoint,\ny)
		}						

		\psline[linewidth=1.2pt,linecolor=colorXAxis]{-}(\xOrig,\yOrig)(\xEndPoint,\yOrig)
																	
		\definecolor{c1}{RGB}{153,0,0}
		\definecolor{c2}{RGB}{122,153,0}
		\definecolor{c3}{RGB}{0,153,61}
		\definecolor{c4}{RGB}{0,61,153}
		\definecolor{c5}{RGB}{122,0,153}
				
		\psset{xunit=\xScale\psunit,yunit=\yScale\psunit,linewidth=0.4pt,dotsize=2.2pt}		
		\psline[linewidth=1.0pt,linecolor=black!90!white,strokeopacity=0.31]{-}(2.25,75)(2.25,424)
		\psline[linewidth=1.0pt,linecolor=black!90!white,strokeopacity=0.31]{-}(0.25,378)(4.25,378)
		\dataplot[xlogBase=10,origin={\xDataOrig,\yDataOrig},showpoints=true,linecolor=c2]{\betaB}
		\dataplot[xlogBase=10,origin={\xDataOrig,\yDataOrig},showpoints=true,linecolor=c1]{\betaA}
		\dataplot[xlogBase=10,origin={\xDataOrig,\yDataOrig},showpoints=true,linecolor=c3]{\betaC}
		\dataplot[xlogBase=10,origin={\xDataOrig,\yDataOrig},showpoints=true,linecolor=c4]{\betaD}
		\dataplot[xlogBase=10,origin={\xDataOrig,\yDataOrig},showpoints=true,linecolor=c5]{\betaE}
		\dataplot[xlogBase=10,origin={\xDataOrig,\yDataOrig},showpoints=true,linecolor=black,linestyle=dotted,dotsep=0.7pt,dotstyle=square*,linewidth=1.2pt,dotsize=3.6pt]{\bound}
		\psset{xunit=\psunit,yunit=\psunit}
		
		\rput[lb](-0.09,4.8){\footnotesize \textbf{\sf Minimum number of measurements for $\ell_1$-$\ell_1$ minimization}}
		\rput[ct](4.5,0.05){\small \textbf{\sf $\mathsf{\beta}$}}
		
		\rput[lb](4.59,3.25){\small \textbf{\sf \color{black!60!white}{$\mathsf{\beta = 1}$}}}
		\rput[rb](8.40,4.08){\small \textbf{\sf \color{black!60!white}{CS bound~\eqref{Eq:PropChandrasekaranBound}}}}
		\rput[rb](8.35,2.52){\small \textbf{\sf $\ell_1$-$\ell_1$ bound \cite{Mota14-CompressedSensingSideInformation}}}
		
	\end{pspicture}
	}
	\vspace{-0.6cm}
	\caption{
		Solid lines: performance of $\ell_1$-$\ell_1$ minimization for~$5$ different matrices versus~$\beta$. Dotted line: bounds in~\cite{Mota14-CompressedSensingSideInformation}, 
		of which~\eqref{Eq:ThmL1L1} gives the particular case~$\beta = 1$ (vertical line). Horizontal line: bound in~\eqref{Eq:PropChandrasekaranBound} for classical CS. 		
	}
	\label{Fig:BetasL1L1}
\end{figure}
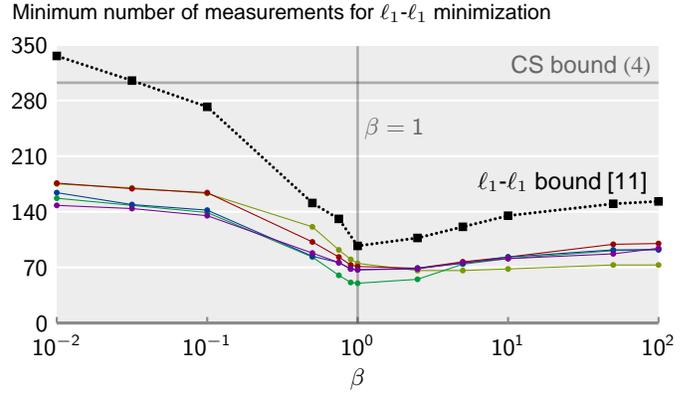

	Our results are illustrated in two types of experiments whose results are shown in Figs.~\ref{Fig:ExperimentalResults} and~\ref{Fig:BetasL1L1}. For \fref{Fig:ExperimentalResults}, we generated a $70$-sparse $x^\star \in \mathbb{R}^{n}$ and a $28$-sparse $w \in \mathbb{R}^n$, where~$n = 1000$; see~\cite{Mota14-CompressedSensingSideInformation} for how they were generated. Although the supports of~$x^\star$ and~$w$ coincided in~$22$ entries (and differed in~$6$), they were significantly different: $\|w - x^\star\|_2/\|x^\star\|_2 \simeq 0.45$ and $\|w - x^\star\|_1/\|x^\star\|_1 \simeq 0.25$. This yielded $\overline{h} = h = 11$, $r = 48$, $\xi = -42$, $v \simeq 103.1$, $q = 76$, and~$K = 1$. Replacing these parameters in the bounds~\eqref{Eq:PropChandrasekaranBound}, \eqref{Eq:ThmL1L1}, and~\eqref{Eq:ThmL1L2Bound}, we have that, for perfect recovery with high probability, standard CS requires at least~$472$ measurements, $\ell_1$-$\ell_1$ minimization requires at least~$136$ measurements, and $\ell_1$-$\ell_2$ minimization requires at least~$666$ measurements, respectively. These values are marked by vertical lines in \fref{Fig:ExperimentalResults}, which shows the experimental performance of standard CS and $\ell_1$-$\ell_1$ and $\ell_1$-$\ell_2$ minimization. Specifically, it depicts the success rate of each scheme as a function of the number of measurements~$m$. For a fixed~$m$, we ran each algorithm~$50$ times, each time for a different (Gaussian) matrix~$A$. The success rate is the number of successful reconstructions over~$50$, the total number of trials. Successful reconstruction here means $\|\hat{x} - x^\star\|_2/\|x^\star\|_2 \leq 10^{-2}$, where~$\hat{x}$ is a solution of~\eqref{Eq:BP} or~\eqref{Eq:BPSideInfoGeneric}. We see that $\ell_1$-$\ell_1$ minimization required less measurements for successful reconstruction than standard CS or $\ell_1$-$\ell_2$ minimization. The performance curves of the last two, in fact, almost coincide, with $\ell_1$-$\ell_2$ (line with triangles) having a slightly sharper phase transition. The figure also shows that, while the $\ell_1$-$\ell_2$ bound~\eqref{Eq:ThmL1L2Bound} can be quite loose, the $\ell_1$-$\ell_1$ bound~\eqref{Eq:ThmL1L1} is quite sharp. 
	In other, unreported experiments, where~$w$ was not sparse, but~$\|w-x^\star\|_2$ was small, a situation apparently very favorable to $\ell_1$-$\ell_2$ minimization, we noticed that $\ell_1$-$\ell_2$ minimization still has a performance similar to CS; of course, in this case, $\ell_1$-$\ell_1$ performs worse than both.

	\fref{Fig:BetasL1L1} considers $\ell_1$-$\ell_1$ minimization only. There, $x^\star \in \mathbb{R}^{500}$ is a $50$-sparse vector and~$w$ was generated such that $\overline{h} = 11$ and $\xi = -30$; see~\cite{Mota14-CompressedSensingSideInformation} for details. We proceeded as follows: we generated a Gaussian matrix $\overline{A} \in \mathbb{R}^{500\times 500}$ and computed $\overline{y} = \overline{A}x^\star$. For a fixed~$\beta$, we solved $\ell_1$-$\ell_1$ minimization using only the first row of~$\overline{A}$ (and of~$\overline{y}$). If the relative error of the solution was larger than~$10^{-2}$, we then used the first two rows of~$\overline{A}$, and so on, until we found a minimal number of measurements~$m(\beta)$ such that $\ell_1$-$\ell_1$ minimization with the first $m(\beta)$ rows of~$\overline{A}$ yielded a relative error smaller than~$10^{-2}$. \fref{Fig:BetasL1L1} shows $m(\beta)$ versus~$\beta$. The solid lines correspond to~$5$ different realizations of $(\overline{A}, \overline{y})$, and the dotted line corresponds to the theoretical $\ell_1$-$\ell_1$ bounds in~\cite{Mota14-CompressedSensingSideInformation}. Note that~\eqref{Eq:ThmL1L1} is the bound for~$\beta = 1$. The plot shows that~$\beta = 1$ minimizes both the theoretical curve and the experimental ones. Also, $\beta = 1$ is the value for which the theoretical bound is the sharpest.
	
\section{Conclusions}

	We integrate side information in CS via $\ell_1$-$\ell_1$ and $\ell_1$-$\ell_2$ minimization and establish bounds on the number of measurements that guarantee successful reconstruction, for Gaussian measurement matrices. Our bound for $\ell_1$-$\ell_1$ minimization is sharp and indicates that if the side information has reasonable quality, $\ell_1$-$\ell_1$ minimization requires much less measurements than both standard CS and $\ell_1$-$\ell_2$ minimization. The underlying geometry of the problem provides an explanation of this phenomenon, and our experimental results also confirm it.

\bibliographystyle{IEEEtran}

\footnotesize
\bibliography{paper}

\end{document}